\documentclass[conference]{IEEEtran}
\IEEEoverridecommandlockouts
\usepackage[english]{babel}
\usepackage{amsthm}
\usepackage{titlesec}
\usepackage[ruled,vlined]{algorithm2e}
\usepackage{algcompatible}
\usepackage{graphicx}
\usepackage{float}
\usepackage{caption}
\usepackage{tikz}
\usepackage{changepage}
\usepackage[utf8]{inputenc}
\usepackage{pgfplots} 
\usepackage{pgfgantt}
\usepackage{pdflscape}
\usepackage[export]{adjustbox}
\pgfplotsset{compat=newest} 
\pgfplotsset{plot coordinates/math parser=false}
\pgfplotsset{compat=1.18}
\usepackage[justification=centering]{caption}
\usepackage{pgfplots}
\usetikzlibrary{spy}

\newtheorem{innercustomgeneric}{\customgenericname}
\providecommand{\customgenericname}{}
\newcommand{\newcustomtheorem}[2]{%
  \newenvironment{#1}[1]
  {%
   \renewcommand\customgenericname{#2}%
   \renewcommand\theinnercustomgeneric{##1}%
   \innercustomgeneric
  }
  {\endinnercustomgeneric}
}

\newcustomtheorem{customthm}{Theorem}
\newcustomtheorem{customlemma}{Lemma}
\newcustomtheorem{customprop}{Proposition}
\newcustomtheorem{customproperty}{Property}
\usepackage{lipsum}
\usepackage{amsmath}
\usepackage{acronym}
\usepackage{amssymb}
\usepackage{mathtools}
\usepackage{url}
\usepackage{graphicx}  
\usepackage{float}  

\newcommand{\yy}{\mathbf{y}}
\newcommand{\xx}{\mathbf{x}}

\newcommand{\bb}{\mathbf{b}}
\newcommand{\zz}{\mathbf{z}}
\newcommand{\hh}{\mathbf{h}}
\newcommand{\nn}{\mathbf{n}}

\newcommand{\ssb}{\mathbf{s}}

\newcommand{\blkdiag}{\textrm{blkdiag}}

\newcommand{\oomega}{\boldsymbol{\omega}}

\newcommand{\cc}{\mathbf{c}}
\newcommand{\BB}{\mathbf{B}}
\newcommand{\UU}{\mathbf{U}}

\newcommand{\GG}{\mathbf{G}}

\newcommand{\HH}{\mathbf{H}}
\newcommand{\ff}{\mathbf{f}}
\newcommand{\VV}{\mathbf{V}}

\newcommand{\QQ}{\mathbf{Q}}
\newcommand{\FF}{\mathbf{F}}

\newcommand{\XX}{\mathbf{X}}
\newcommand{\RR}{\mathbf{R}}

\newcommand{\Tr}{\text{Tr}}
\newcommand{\diag}{\text{diag}}

\newcommand{\hermit}{\mathsf{H}}
\newcommand{\ww}{\mathbf{w}}

\usepackage{accents}

\newcommand{\TT}{\mathbf{T}}

\newcommand{\AAb}{\mathbf{A}}

\newcommand{\norm}[1]{\left\lVert#1\right\rVert}

\newcommand{\TTheta}{\boldsymbol{\Theta}}
\newcommand{\ttheta}{\boldsymbol{\theta}}
\newcommand{\ggc}{\mathbf{g}}
\newcommand{\vecc}{\text{vec}}

\newcommand{\EE}{\mathbf{E}}

\newcommand{\atx}{\mathbf{a}}

\acrodef{SISO}[SISO]{single-input single-output}
\acrodef{AP}[AP]{access point}
\acrodef{UE}[UE]{user equipment}
\acrodef{ULA}[ULA]{uniform linear array}
\acrodef{CPU}[CPU]{central processing unit}
\acrodef{FPP}[FPP]{Feasible-point pursuit}
\acrodef{LoS}[LoS]{line-of-sight}
\acrodef{NLoS}[NLoS]{non-line-of-sight}
\acrodef{RCS}[RCS]{radar cross section}
\acrodef{AoD}[AoD]{angle of departure}
\acrodef{AoA}[AoA]{angle of arrival}
\acrodef{CRB}[CRB]{Cramer-Rao bound}
\acrodef{FIM}[FIM]{Fisher information matrix}
\acrodef{AN}[AN]{artificial noise}
\acrodef{SINR}[SINR]{signal-to-interference-plus-noise ratio}
\acrodef{SNR}[SNR]{signal-to-noise ratio}
\acrodef{QoS}[QoS]{quality of service}
\acrodef{SDR}[SDR]{semi-definite relaxation}
\acrodef{SDP}[SDP]{semi-definite relaxation}
\acrodef{ISAC}[ISAC]{integrated sensing and communications}
\acrodef{PLS}[PLS]{physical layer security}
\acrodef{SIC}[SIC]{successive interference cancellation}
\acrodef{CSI}[CSI]{channel state information}
\acrodef{MUI}[MUI]{multi-user interference}
\acrodef{RIS}[RIS]{reconfigurable intelligent surface}
\acrodef{AO}[AO]{alternating optimization}
\acrodef{SIMO}[SIMO]{Single Input Multiple Output}
\acrodef{MISO}[MISO]{multiple-intput single output}
\acrodef{MIMO}[MIMO]{multiple-input multiple-output}
\acrodef{MU}{multi-user}
\acrodef{BS}[BS]{base station}
\acrodef{CEE}[CEE]{channel estimation error}
\acrodef{CCP}[CCP]{convex-concave procedure}
\acrodef{MRT}[MRT]{Local maximum-ratio transmission}
\acrodef{MM}[MM]{minorization-maximization}
\acrodef{PSD}[PSD]{positive semi-definite}
\acrodef{RZF}[RZF]{regularized zero forcing}
\acrodef{CRZF}[CRZF]{Centralized regularized zero forcing}
\acrodef{LRZF}[LRZF]{Local regularized zero forcing}
\acrodef{LPZF}[LPZF]{Local Partial zero forcing}
\acrodef{LZF}[LZF]{Local zero forcing}
\acrodef{GLRT}[GLRT]{generalized likelihood ratio test}
\acrodef{CDF}[CDF]{cumulative distribution function}
\acrodef{SOC}[SOC]{second order cone}
\acrodef{SCNR}[SCNR]{signal-to-clutter-and-noise ratio}
\acrodef{CNR}[CNR]{clutter-to-noise ratio}
\acrodef{UPA}[UPA]{uniform linear array}
\begin{document}

\title{Clutter-Aware Target Detection for ISAC in a Millimeter-Wave Cell-Free Massive MIMO System }

\author{Steven~Rivetti$^\dagger$,
Özlem~Tu$\Breve{\text{g}}$fe~Demir$^*$,
Emil~Björnson$^\dagger$, 
        Mikael~Skoglund$^\dagger$ \\
        
       {\small$^\dagger$School of Electrical Engineering and Computer Science (EECS),
        KTH Royal Institute of Technology, Sweden} \\
        
        {\small$^*$Department of Electrical-Electronics Engineering, TOBB University of Economics and Technology, Ankara, Türkiye}
        
       \thanks{
This work was supported by the SUCCESS project (FUS21-0026), funded by the Swedish Foundation for Strategic Research. \"O. T. Demir was supported by 2232-B International Fellowship for Early Stage Researchers Programme funded by the Scientific and Technological Research Council of T\"urkiye.}}

\maketitle

\begin{abstract}
In this paper, we investigate the performance of an integrated sensing and communication (ISAC) system within a cell-free massive multiple-input multiple-output (MIMO) system. Each access point (AP) operates in the millimeter-wave (mmWave) frequency band. The APs jointly serve the user equipments (UEs) in the downlink while simultaneously detecting a target through dedicated sensing beams directed toward a reconfigurable intelligent surface (RIS). Although the AP-RIS, RIS-target, and AP-target channels have both line-of-sight (LoS) and non-line-of-sight (NLoS) parts, only knowledge of the LoS paths is assumed to be available. A key contribution of this study is the consideration of clutter, which degrades the target detection performance if not handled.
We propose an algorithm to alternatively optimize the transmit power allocation and the RIS phase-shift matrix, maximizing the target signal-to-clutter-plus-noise ratio (SCNR) while ensuring a minimum signal-to-interference-plus-noise ratio (SINR) for the UEs. Numerical results demonstrate that exploiting clutter subspace significantly enhances detection probability, particularly at high clutter-to-noise ratios, and reveal that an increased number of transmit side clusters impairs detection performance. Finally, we highlight the performance gains achieved using a dedicated sensing stream.

\end{abstract}

\begin{IEEEkeywords}
RIS, ISAC, cell-free massive MIMO, mmWave.
\end{IEEEkeywords}

\section{Introduction}
Cell-free massive \ac{MIMO} is a user-centric network infrastructure where a set of distributed \acp{AP} serve multiple \acp{UE} on the same time-frequency resources using joint processing techniques \cite{demir2021foundations}. Over nearly a decade of research, it has been shown that cell-free massive MIMO improves both spectral and energy efficiency by providing macro-diversity and enhanced interference management compared to traditional cellular setups. As a result, it has become one of the key technology components considered for sixth-generation (6G) wireless networks \cite{ngo2024ultradense}. Recently, cell-free massive MIMO architecture has also demonstrated potential benefits in \ac{ISAC} \cite{behdad2024multi,demirhan2023cell}. ISAC allows for the efficient use of hardware and spectral resources through integration and coordination gains, compared to separate communication and sensing architectures \cite{lu2024integrated}. Cell-free massive MIMO facilitates multi-static sensing through a central processing unit (CPU), to which the APs are connected via fronthaul links. Beyond cell-free massive MIMO and ISAC, another key 6G technology is \ac{RIS}-aided communications. 
RISs have emerged as a promising technology in ISAC applications, enhancing the available spatial degrees of freedom.
In \cite{liu2022joint}, the sum rate of communication UEs is maximized under a worst-case sensing signal-to-noise ratio (SNR) constraint. 
An often overlooked aspect in ISAC research is the effect of clutter on target detection, which can severely degrade system performance if not properly addressed. Recent studies, such as \cite{demir2024ris}, illustrate the impact of clutter in monostatic ISAC settings, underscoring the significance of clutter awareness at the receiving \acp{AP} for reliable target detection.
\begin{figure}[t]
\begin{center}
   \resizebox{0.45\textwidth}{!}{
    \input{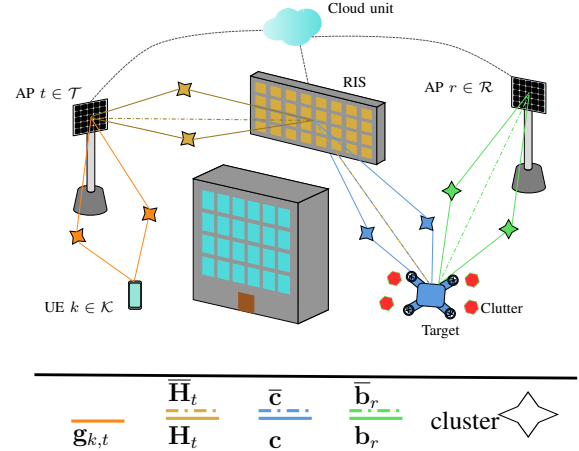}}

	 \caption{Depiction of a cell-free massive MIMO ISAC network performing multi-static sensing. } \label{general scheme}
		\end{center}
  \vspace{-6mm}
\end{figure}
The contributions of this paper are outlined as follows
\begin{itemize}
    \item we investigate the ability of a RIS to improve target detection under transmitter-target obstruction. The need for precise \ac{CSI} for target detection and the requirement for sensing-specific streams are explored within a millimeter-wave (mmWave) channel architecture.
    \item A \ac{SCNR}-maximizing \ac{AO} algorithm is proposed, where the system assumes a \ac{LoS} structure for the sensing links and optimizes the transmit power allocation and the RIS phase-shift matrix
    \item numerical simulations reveal that clutter awareness and dedicated sensing streams significantly enhance the detection probability.
\end{itemize}
\vspace{-5mm}
\section{System Model}
\vspace{-1mm}
We consider the \ac{RIS}-assisted cell-free massive \ac{MIMO} \ac{ISAC} network shown in Fig. \ref{general scheme}. A set $\mathcal{T}$ of $T$ \acp{AP}, each equipped with $M$ transmit antennas, serves $K$ single-antenna \acp{UE} in the downlink while sensing the potential presence of a target at a known position.
Sensing is carried out over $\tau$ symbols belonging to the same channel coherence block.
A second set $\mathcal{R}$ of $R$ APs collects the target echoes and performs target detection.
The choice of geographically separated transmitting and receiving APs is motivated by the implementation challenges of full duplex operation, especially in terms of self-interference \cite{he2024research}, and the interest in this setup for cell-free massive MIMO communications.
We assume that the direct channel between the APs in $\mathcal{T}$ and the target is blocked and, thus, the formers reach the target through a  \ac{RIS} equipped with $N$ reflective elements, whose phase shifts are described by the vector $\ttheta=[\theta_1,\dots, \theta_N]^\top$, where $|\theta_n|=1$ for $n=1,\dots, N$.
Each AP $t \in \mathcal{T}$ is connected to the RIS via the channel $\HH_{t} \in \mathbb{C}^{N \times M}$, the reflected path connecting the RIS to the target is denoted by $\cc\in \mathbb{C}^{N} $ and the  UEs are connected to the APs in $\mathcal{T}$ via the channel $\{\ggc_{k,t}\}_{\forall k,t} \in \mathbb{C}^{M} $.
The receiving APs in $\mathcal{R}$ are connected to the target via the channels $\{\bb_r\}_{\forall r \in \mathcal{R}}\in \mathbb{C}^M$.
\vspace{-1mm}
\subsection{Channel models}
\vspace{-1mm}
 Each of the previously mentioned channels is modelled in accordance with \cite{zhang2019hybrid}, which accurately captures the spatial sparsity that characterizes mmWave channels. Each channel consists of contributions from $C_i$ scattering clusters, i.e.,
\begin{align}
    &\bb_{r} =\underbrace{\sqrt{\beta^2_{0,r}}\hspace{1mm} \atx_M(\oomega_{0,r}^{\textrm{RX}})}_{\overline{\bb}_r}+\sqrt{\frac{1 }{C_1}}\sum_{n=1}^{C_1}
\alpha_{n,r}\hspace{0.5mm}\atx_M(\oomega_{n,r}^{\textrm{RX}}),\\
    &\cc =\underbrace{\sqrt{\beta_{0}^2}\hspace{1mm}  \atx_N(\oomega_{0}^{\textrm{Tg}})}_{\overline{\cc}} +  \sqrt{\frac{1}{C_2}}\sum_{n=1}^{C_2}
\alpha_{n}\hspace{0.5mm}\atx_N(\oomega_{n}^{\textrm{Tg}}),\\
    & \ggc_{k,t} =\sqrt{\frac{1}{C_3}}\sum_{n=1}^{C_3}
\alpha_{n,k,t}\hspace{0.5mm}\atx_M(\oomega_{n,k,t}^{\textrm{UE}}),\\
    &\HH_t =\overbrace{\sqrt{\beta_{0,t}^2}\hspace{1mm} \atx_N(\oomega_{0,t}^{\textrm{RIS}})\atx_M^\top(\oomega_{0,t}^{\textrm{TX}})}^{\overline{\HH}_t} \nonumber\\
    &\quad + \sqrt{\frac{1}{C_4}}\sum_{n=1}^{C_4}
\alpha_{n,t}\hspace{0.5mm}\atx_N(\oomega_{n,t}^{\textrm{RIS}})\atx_M^\top(\oomega_{n,t}^{\textrm{TX}}), 
\end{align}
where the overline superscript indicates the LoS paths and the other terms correspond to the NLoS parts. The coefficients $\beta_{0,r}^2$, $\beta_{0}^2$, and $\beta_{0,t}^2$ are the corresponding LoS channel gains.
We now remove the additional subscripts and superscripts except for $n$ as the following consideration applies regardless of the additional subscripts. Here, $\alpha_n$ is the complex channel gain of the $n$-th cluster. We assume each cluster scatters a sufficient number of rays such that  $\alpha_n \sim \mathcal{CN}(0,\beta_n^2)$  and the fading is mutually independent.
The coefficients $\beta_n^2$ represent the path loss associated with the respective clusters.
The geometry of the $n$-th cluster is described by $\oomega_n =[\psi_n,\phi_{n}]$, where $\psi_n$   is the azimuth angle  and $\phi_n$ is the elevation angle. 
Each AP is equipped with a  uniform planar square array with a half-wavelength vertical and horizontal inter-element spacing such that the entries of $M$-dimensional steering vector are defined as $   [\atx_M(\oomega_n)]_{m_hm_v}=
e^{-j\pi( m_h\sin(\psi_n)\cos(\phi_n) + m_v\sin(\phi_n))}$,for $m_h,m_v =0, \dots,\sqrt{M}-1$. We assume that perfect CSI regarding $\{\ggc_{k,t}\}_{\forall k,t}$  is available at the APs, as this is easily obtainable through standard estimation techniques. As for the sensing channels, the transmit APs know the position of the RIS and the target and, thus, the corresponding LoS channels.
\subsection{Communication observation model}
We assume that the APs are equipped with fully digital beamforming hardware: $\ff_{l,t}$ is the precoding vector for stream $l$ and AP $t$, and $\FF_t = [\ff_{1,t},\ldots,\ff_{L,t}] \in \mathbb{C}^{M \times L}$, where $L=K+1$. Here, we allocate one beam to each UE and an additional one, denoted by the index $L$, for sensing purposes, bringing the total number of digital beams to $L=K+1$. This choice is motivated by the fact that the UEs and target typically are at very different positions, and catering solely to the communication UEs might drastically reduce the system's sensing performance.

The received downlink signal at \ac{UE} $k$ during timeslot $\iota$ is defined as  
\begin{align}
    y_k[\iota] &= \sum_{t \in \mathcal{T}}\ggc_{k,t}^\hermit \ssb_t[\iota]
    + w_k[\iota] =\sum_{t \in \mathcal{T}}\Big(\underbrace{\sqrt{\rho_{k}}\ggc_{k,t}^\hermit \ff_{k,t}x_{k}[\iota]}_{\text{desired signal}} \\
    &\hspace{-8mm}+ \underbrace{\sum_{k' \in \mathcal{K} \setminus \{ k \}}\sqrt{\rho_{k'}}\ggc_{k,t}^\hermit \ff_{k',t}x_{k'}[\iota]}_{\text{multi-user interf. (MUI)}}  +\underbrace{\sqrt{\rho_{L}}\ggc_{k,t}^\hermit \ff_{L,t}x_{L}[\iota]}_{\text{sensing interf.}} \Big)+ w_k[\iota],\nonumber
\end{align}
with $\ssb_t[\iota] =\FF_{t}\XX[\iota]\boldsymbol{\rho}$, where the matrix $\XX[\iota]=\diag(x_1[\iota],\ldots,x_L[\iota])$ is a diagonal matrix containing the downlink communication symbols and sensing symbol transmitted during timeslot $\iota$, where $\mathbb{E}\{|x_k[\iota]|^2\}=1$.
The vector $\boldsymbol{\rho}=[\sqrt{\rho_1},\dots,\sqrt{\rho_L}]^\top$ contains the square root of the power allocated to each beam.
 Finally, $w_k[\iota]\sim \mathcal{CN}(0,\sigma_k^2)$.
In line with the centralized operation of cell-free massive MIMO, the communication precoding vectors are based on the transmit APs' CSI \cite{behdad2024multi} while the sensing precoder is based on the assumed LoS structure of the sensing channels and the known target location. To this end, we
define $\ggc_k=[\ggc_{k,1}^\top,\dots,\ggc_{k,T}^\top]^\top \in \mathbb{C}^{MT}$ and $\ff_k=[\ff_{k,1}^\top,\dots,\ff_{k,T}^\top]^\top \in \mathbb{C}^{MT}$. The communication precoding vectors for each UE are chosen using \ac{RZF} as $\ff_{k}=\overline{\ff}_{k}/\norm{\overline{\ff}_{k}}$, where
\begin{align}\label{LRZF}
 &\overline{\ff}_{k} =\left( \sum_{i \in \mathcal{K}}\ggc_{i}\ggc_{i}^\hermit + \lambda\mathbf{I}_{MT} \right)^{-1}\ggc_{k}
\end{align}
where $\lambda$ is a regularization parameter.
Let us define the matrix $\GG=[\ggc_{1},\dots,\ggc_{K}] \in \mathbb{C}^{MT \times K}$ with all the UE channels. Then the sensing stream precoder is defined as $\ff_{L,t}=\overline{\ff}_{L,t}/\norm{\overline{\ff}_{L,t}}$
\begin{align}
     \overline{\ff}_{L,t}=\left( \mathbf{I}_{MT} -\GG \left( \GG^\hermit \GG  \right)^\dagger \GG^\hermit 
  \right)\overline{\hh}
\end{align}
where $\overline{\hh}=[\overline{\hh}_{1}^\top,\dots,\overline{\hh}_{T}^\top]^\top$ and $\overline{\hh}_{t}=\overline{\HH}_t^\hermit\TTheta\overline{\cc}$ with $\TTheta= \diag(\ttheta)$.
We have projected $\overline{\hh}$  onto the null space of the communication UEs' channels to nullify the interference caused to the UEs.
The communication performance is represented by the \acp{UE}' spectral efficiency (SE), defined as ${\rm{SE}}_k = \log_2(1+ {\rm{SINR}}_k) $: under the assumption of perfect \ac{CSI} on the UEs channels, the \ac{SINR} is defined as 
\begin{align}
    {\rm{SINR}}_k = \frac{  |{\ggc}_{k}^\hermit\ff_{k} \sqrt{\rho_{k}}|^2 }{\sum_{k' \in \mathcal{K}/k}|{\ggc}_{k}^\hermit\ff_{k'} \sqrt{\rho_{k'}}|^2 + |{\ggc}_{k}^\hermit\ff_{L} \sqrt{\rho_{L}}|^2 + \sigma_k^2}.
\end{align}
\subsection{Sensing observation model}
The two-way sensing channel between transmit  AP $t$ and receive AP $r$ is defined as $ \EE_{t,r}= \bb_r\hh_{t}^\hermit \triangleq\bb_r\left( \HH_t^\hermit\TTheta\cc\right)^\hermit.$
The  sensing observation of  AP $r$ at timeslot $\iota$  is 
\begin{align} \label{eq:sensingobserv}
    \yy_r[\iota] = \sum_{t\in \mathcal{T}} \xi_{t,r} \EE_{t,r} \ssb_t[\iota] + \zz_r[\iota] + \nn_r[\iota],
    \vspace{-2mm}
\end{align}
where $\xi_{t,r}$ is the \ac{RCS} associated with the sensing path between transmit AP $t$ and receive AP $r$. We adopt the Swerling-I model, meaning that the RCS assumes only one value during the transmission and $\xi_{t,r} \sim \mathcal{CN}(0,\delta_{t,r}^2)$. We assume that $\{\delta_{t,r}^2\}_{\forall t,r}$ are known and that RCSs belonging to different transmit-receiving pairs are statistically uncorrelated, that is $ \mathbb{E}\left\{\xi_{a,r}\xi_{a',r'}^*\right\} \neq 0 \Leftrightarrow a=a',~r=r'$.
In \eqref{eq:sensingobserv}, $\nn_r[\iota]$ is the receiver noise of AP $r$, assumed to have independent $\mathcal{CN}(0,\sigma^2)$ entries.
The vector $\zz_r[\iota]$ represents the clutter as observed from AP $r$. The clutter vector is modelled as $\zz_r[\iota]\sim \mathcal{CN}(\boldsymbol{0},\delta_{\zz}^2\RR_r)$, where $\delta_{\zz}^2$ is the power of the clutter cross-section, which also includes the path loss effects, thus defining the \ac{CNR} as $\textrm{CNR}=\delta_{\zz}^2/\sigma^2$. The normalized clutter covariance matrix, with a trace equal to $M$, is denoted by $\RR_r$. This matrix is unknown, but its eigenspace is assumed to be known later in the detector design.
For the sake of mathematical tractability, we assume that the clutter observations belonging to different APs and/or different timeslots are statistically uncorrelated.
Under the Los assumption for the channels involved in sensing, the target's \ac{SCNR} is defined as
\begin{align}\label{SNR og}
    &\overline{\textrm{SCNR}}=
    \frac{\mathbb\sum_{\iota =1}^\tau\mathbb\sum_{r \in \mathcal{R}} \sum_{a\in \mathcal{T}}\delta_{a,r}^2
     \norm{\overline{\EE}_{a,r} \ssb_{a}[\iota]}^2}{ R\tau M\left( \sigma^2 + \delta_{\zz}^2\right)},
\end{align}
where $\overline{\EE}_{a,r}=\overline{\bb}_r\left( \overline{\HH}_a^\hermit\TTheta\overline{\cc}\right)^\hermit$.

\section{Sensing performance optimization}
The system aims at jointly maximizing
the sensing SCNR \eqref{SNR og} while guaranteeing a minimum \ac{SINR} to all the communication \acp{UE}. 
The optimization problem is stated as follows:
\begin{subequations}
\begin{align}
     \underset{\ttheta,\boldsymbol{\rho}}{\mathrm{maximize}} \,\, &~  
     \overline{\textrm{SCNR}}\\
     \text{subject to} ~& \textrm{SINR}_k  \geq \gamma_k,\quad \forall k \in \mathcal{K}, \label{sinr og} \\
     & \sum_{l=1}^L\norm{\ff_{l,t}}^2\rho_l \leq P_t,\quad \forall t \in \mathcal{T}, \label{power og} \\
     & |\theta_n| = 1, \quad n=1,\dots, N, 
\end{align}
\end{subequations}
where $\gamma_k$ is the SINR threshold of UE $k$ and  $P_t$ is the available transmit power of the $t$-th \ac{AP}.
This problem is non-convex, mainly due to the coupling between the optimization variables and the non-convex unitary modulus constraints.
In this section, we will devise an \ac{AO} algorithm, which switches between optimizing the power allocation policy and the RIS phase shifts until convergence is achieved.
\subsection{Power allocation optimization}\label{precoder subs}
In this subsection, we introduce the first subproblem of the \ac{AO} algorithm, where we optimize the power allocation for a fixed RIS configuration. We first reformulate \eqref{SNR og} as $\overline{\textrm{SCNR}}=\boldsymbol{\rho}^\top \Re( \AAb ) \boldsymbol{\rho}$
where 
\begin{align}
    \AAb=\frac{\sum_{\iota=1}^\tau\mathbb\sum_{r \in \mathcal{R}} \sum_{a\in \mathcal{T}}\delta_{a,r}^2\XX^\hermit[\iota]\FF_a^\hermit\overline{\EE}_{a,r}^\hermit \overline{\EE}_{a,r}\FF_a\XX[\iota]}{R\tau M\left( \sigma^2 + \delta_{\zz}^2\right)}.
\end{align}
The constraints \eqref{sinr og} and \eqref{power og} can be reformulated as \ac{SOC} constraints in terms of $\boldsymbol{\rho}$.
The first subproblem of the \ac{AO} SCNR-maximizing algorithm is  defined as
\vspace{-0.2mm}
\begin{subequations}\label{P1}
\begin{align}
    \underset{\boldsymbol{\rho}}{\mathrm{maximize}} \,\, &~  
    \boldsymbol{\rho}^\top \Re( \AAb ) \boldsymbol{\rho}
    \\
    \text{subject to} ~ &  
        \Big[|\ggc_k^\hermit\ff_1|\sqrt{\rho_1} \quad \dots \quad \underbrace{0}_k \quad \dots \quad|\ggc_k^\hermit\ff_L|\sqrt{\rho_L} \nonumber
        \\
        &\sigma_k \Big] 
    \leq (|\ggc_k^\hermit\ff_k|\sqrt{\rho_k})/{\sqrt{\gamma_k}},~\forall k\in \mathcal{K}\\
   & \norm{\widetilde{\FF}_t\boldsymbol{\rho}} \leq \sqrt{P_t},~\forall t\in \mathcal{T}
\end{align}
\end{subequations}
where $\widetilde{\FF}_t=\diag\left(\norm{\ff_{1,t}},\dots, \norm{\ff_{L,t}} \right)$.
This problem can be solved using the convex-concave procedure outlined in \cite{behdad2024multi}.

\subsection{RIS phase-shift optimization}\label{RIS subs}
We now move on to the second part of the \ac{AO}, where we optimize $\ttheta$ when the power variables are fixed. 
The SCNR can be reformulated as $\overline{\textrm{SCNR}}=\ttheta^\hermit\QQ\ttheta$,
where 
\begin{align}
    \QQ=&\frac{1}{R\tau M\left( \sigma^2 + \delta_{\zz}^2\right)}\sum_{\iota=1}^\tau\sum_{r \in \mathcal{R}} \sum_{a\in \mathcal{T}} \Big( \delta_{a,r}^2\overline{\cc}\overline{\bb}_r^\hermit\overline{\bb}_r \overline{\cc}^\hermit\Big)^\top  \nonumber \\ 
&\quad \quad \odot  \left(\overline{\HH}_{a}\ssb_a[\iota]
    \ssb_a^\hermit[\iota] \overline{\HH}_{a}^\hermit \right) .
\end{align}
We have used $\Tr(\AAb\XX\BB\XX^\hermit)=\vecc^\hermit(\XX)\left(\BB^\top \otimes \AAb\right) \vecc(\XX)$ and $\text{\normalfont vec}^\hermit(\diag(\xx)) ( \AAb \otimes \BB ) \text{\normalfont vec}(\diag(\xx)) = \xx^\hermit (\AAb \odot \BB ) \xx$ from \cite{horn2012matrix}. The RIS phase-shift optimization problem under consideration is still non-convex: We then employ the \ac{MM} algorithm to maximize a convex lower bound on the objective function \cite{mehanna2014feasible} around the local point $\ttheta^{(s)}$.
We can now define the second subproblem of the AO optimization as
\begin{subequations}\label{P2}
\begin{align}
 \underset{\ttheta,\alpha \geq 0}{\textrm{maximize}}&~ \alpha   \\
     \text{subject to} ~&\alpha \leq  2\Re\left(\ttheta^{(s)\hermit}\QQ^{(+)}\ttheta\right) + \ttheta^\hermit\QQ^{(-)}\ttheta  \\
&\quad \quad -\ttheta^{(s)\hermit}\QQ^{(+)}\ttheta^{(s)}, \nonumber\\
& |\theta_n| \leq 1 ,\quad n=1,\dots, N,
 \end{align}
 \end{subequations}
 where $\alpha$ is an auxiliary variable and $\QQ^{(+)}$ and $\QQ^{(-)}$ represent the positive and negative semidefinite parts of $\QQ$.
 Note that we have relaxed the unit modulus constraint, as it is observed that the solution satisfies this constraint with equality anyway.
This is a convex problem and can be solved at each iteration until convergence is achieved.
The steps of the AO algorithm are outlined in Algorithm~1.

\begin{algorithm} [h!]
  \caption{ SCNR maximizing \ac{AO} algorithm}
  \begin{algorithmic}[1]
    \STATE \textbf{Initialize:} Generate $\ttheta^{(0)}$ randomly, $v \leftarrow 0$
    \REPEAT
        \STATE Compute the precoding codebooks with $\ttheta = \ttheta^{(v)}$
        \STATE Obtain $\boldsymbol{\rho}^{(v+1)}$ by solving \eqref{P1}
         \STATE Obtain $\ttheta^{(v+1)}$ by solving  \eqref{P2} iteratively
         \STATE $v \leftarrow v+1$
    \UNTIL{Convergence}
    \STATE \textbf{Output:} $\ttheta^\text{opt},\boldsymbol{\rho}^\text{opt}$
  \end{algorithmic}
\end{algorithm}
\vspace{-5mm}
 \section{GLRT Detector}
 One of the main metrics used to assess the sensing performance is the probability of correct detection, implemented through a \ac{GLRT} by extending the detector in  \cite{demir2024ris} to a cell-free massive MIMO system based on the optimization procedure developed in the previous section.
 We operate under the assumption that the transmit signals are known by the receiving APs, thanks to the cell-free architecture.
We define the vector containing the combined sensing observation of all receiving APs during timeslot $\iota$ as 
 \begin{align*}\label{yy}
    & \yy[\iota]  =[\yy_1^\top[\iota], \dots,\yy_R^\top[\iota]]^\top= \VV[\iota]\boldsymbol{\xi} + \UU\mathbf{w}[\iota] + \nn[\iota],\\
&\VV[\iota]=\blkdiag(\VV_1[\iota],\dots,\VV_R[\iota])\in \mathbb{C}^{MR \times TR},\\
&\VV_r[\iota] = \left[ \EE_{1,r}\ssb_1[\iota],\dots, \EE_{T,r}\ssb_T[\iota]  \right]\in \mathbb{C}^{M \times T},\\
&\nn[\iota]=[\nn_1^\top[\iota], \dots,\nn_R^\top[\iota]]^\top ,~\boldsymbol{\xi}=[\xi_{1,1},\dots,\xi_{T,R}]^\top, \\
&\UU=\blkdiag\left( \UU_1 , \dots , \UU_R \right) , \ww[\iota] = [\ww_1^\top[\iota], \dots, \ww_R^\top[\iota] ]^\top.
 \end{align*}
The clutter observed by AP $r$ is assumed to exist in a low-rank subspace \cite{breloy2015clutter} spanned by the columns of the semi-unitary matrix $\UU_r \in \mathbb{C}^{M \times i}$, with $i$ being the subspace dimension, obtained from the eigenspace of  $\delta_{\zz}^2\RR_r$.
AP $r$'s clutter is then expressed as $\UU_r\mathbf{w}_r[\iota]$, where the distribution of $\mathbf{w}_r[\iota]$ is unknown. The previous observations are generated using the true channels; however, the system does not have perfect CSI on the sensing channels and thus assumes a LoS structure. Therefore, the binary hypotheses  being tested are 
\begin{subequations}\label{detector}
  \begin{align}
    & \mathcal{H}_0 : \yy[\iota]  = \overline{\VV}[\iota]\boldsymbol{\xi} + \UU\mathbf{w}[\iota] + \nn[\iota],~\iota=1,\dots, \tau,\\
    & \mathcal{H}_1 : \yy[\iota] =\UU\mathbf{w}[\iota] + \nn[\iota],~\iota=1,\dots, \tau,
\end{align}  
\end{subequations}
where $\overline{\VV}$ is computed as $\VV$ by using $\overline{\EE}_{t,r}$.
The adoption of a GLRT method is motivated by the unknown distribution of $\ww[\iota]$, however, we know that $\yy_{\mathcal{H}_0}[\iota] -\UU\mathbf{w}[\iota]\sim \mathcal{CN}(\mathbf{0},\sigma^2\mathbf{I}_{MR})$ and $\yy_{\mathcal{H}_1}[\iota] -\UU\mathbf{w}[\iota]\sim \mathcal{CN}(\mathbf{0},\RR[\iota]+\sigma^2\mathbf{I}_{MR})$, where $\RR[\iota]=\mathbb{E}\left\{ \overline{\VV}[\iota]\boldsymbol{\xi} \boldsymbol{\xi}^\hermit  \overline{\VV}^\hermit[\iota]\right\} =\overline{\VV}[\iota] \diag(\delta^2_{1,1},\dots,\delta^2_{T,R})\overline{\VV}^\hermit[\iota] $.
Following the derivation in \cite{demir2024ris}, the GLRT test can be rewritten as 
\begin{align}
    &\sum_{\iota=1}^\tau \yy^\hermit[\iota]\Big(\boldsymbol{\Xi}[\iota]\UU\Big( \UU^\hermit \boldsymbol{\Xi}[\iota] \UU\Big)^{-1}\UU^\hermit\boldsymbol{\Xi}[\iota] \nonumber  + \\
    &\sigma^{-2}(\mathbf{I}_{MR} - 
    \UU\UU^\hermit)-
    \boldsymbol{\Xi}[\iota]\Big) \yy[\iota]\lessgtr^{\mathcal{H}_0}_{\mathcal{H}_1}  \widetilde{\lambda}_d
\end{align}
where $\boldsymbol{\Xi}[\iota]=(\RR[\iota]+\sigma^2\mathbf{I}_{MR})^{-1}$.  $\widetilde{\lambda}_d=\ln(\lambda_d) - \ln\left( \det\left(\sigma^2\mathbf{I}_{MR}\right)^\tau/{\prod_{\iota=1}^\tau\det\left(\RR[\iota]+\sigma^2\mathbf{I}_{MR}\right)}\right)$ is a revised threshold whose value should be chosen according to the desired false alarm probability.

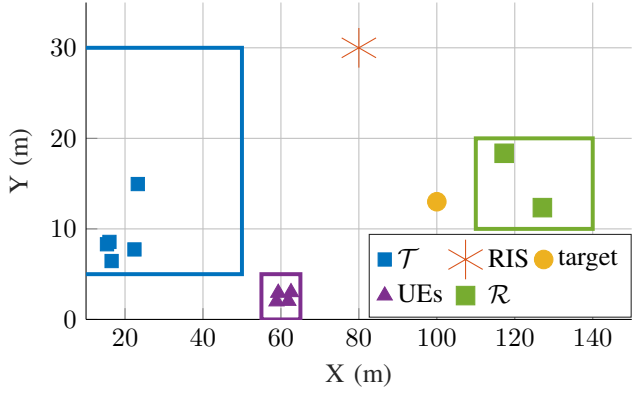
\begin{figure}[t]
\begin{center}
%
%
\definecolor{mycolor1}{rgb}{0.00000,0.44700,0.74100}%
\definecolor{mycolor2}{rgb}{0.00000,0.44706,0.74118}%
\definecolor{mycolor3}{rgb}{0.85000,0.32500,0.09800}%
\definecolor{mycolor4}{rgb}{0.85098,0.32549,0.09804}%
\definecolor{mycolor5}{rgb}{0.92900,0.69400,0.12500}%
\definecolor{mycolor6}{rgb}{0.92941,0.69412,0.12549}%
\definecolor{mycolor7}{rgb}{0.49400,0.18400,0.55600}%
\definecolor{mycolor8}{rgb}{0.49412,0.18431,0.55686}%
\definecolor{mycolor9}{rgb}{0.46600,0.67400,0.18800}%
\definecolor{mycolor10}{rgb}{0.46667,0.67451,0.18824}%
\begin{tikzpicture}

\begin{axis}[%
width=0.4\textwidth,
height=1.26in,
at={(0.758in,0.481in)},
scale only axis,
xmin=10,
xmax=150,
xlabel style={font=\color{white!15!black}},
xlabel={x (m)},
ymin=0,
ymax=35,
legend columns =3,
ylabel style={font=\color{white!15!black}},
ylabel={y (m)},
axis background/.style={fill=white},
axis x line*=bottom,
axis y line*=left,
xmajorgrids,
ymajorgrids,
legend style={at={(1,0.97)},legend cell align=left, align=left, draw=white!15!black}
]
\addplot [color=mycolor1, only marks, mark size=2.5pt, mark=square*, mark options={solid, fill=mycolor2, draw=mycolor2}]
  table[row sep=crcr]{%
16.0031898144935	8.56257641208275\\
23.2748535731924	14.9527532186242\\
15.3735336691483	8.30312360243299\\
22.4186708775931	7.73187682260489\\
16.5559327331883	6.45002697590095\\
};
\addlegendentry{$\mathcal{T}$}

\draw[line width=1.5pt, draw=mycolor2] (axis cs:0,5) rectangle (axis cs:50,30);
\addplot [color=mycolor3, only marks, mark size=7.5pt, mark=pentagon*, mark options={solid, fill=mycolor3, draw=mycolor3}]
  table[row sep=crcr]{%
80	30\\
};
\addlegendentry{RIS}

\addplot [color=mycolor5, only marks, mark size=3.5pt, mark=*, mark options={solid, fill=mycolor6, draw=mycolor6}]
  table[row sep=crcr]{%
100	13\\
};
\addlegendentry{target}

\addplot [color=mycolor7, only marks, mark size=3.3pt, mark=triangle*, mark options={solid, fill=mycolor8, draw=mycolor8}]
  table[row sep=crcr]{%
58.86235039062606	1.97865684840299\\
59.31111627487221	2.93361860386109\\
61.90784648009122	2.03840081846848\\
62.58582251418772	3 .12809588835559\\
};
\addlegendentry{UEs}

\draw[line width=1.5pt, draw=mycolor8] (axis cs:55,0) rectangle (axis cs:65,5);
\addplot [color=mycolor9, only marks, mark size=3.5pt, mark=square*, mark options={solid, fill=mycolor10, draw=mycolor10}]
  table[row sep=crcr]{%
117.275566779717	18.3504194801815\\
127.083003861937	12.3474282867107\\
};
\addlegendentry{$\mathcal{R}$}

\draw[line width=1.5pt, draw=mycolor10] (axis cs:110,10) rectangle (axis cs:140,20);
\end{axis}
\end{tikzpicture}%
      \vspace{-2mm}
	 \caption{2D locations in the simulated scenario.}  \label{simulation scen}
		\end{center}
  \vspace{-8mm}
\end{figure}

\section{Numerical Results}
A  two-dimensional view of the simulated scenario is provided in Fig.~\ref{simulation scen}.
The locations of the APs in $\mathcal{T}$, the APs in $\mathcal{R}$, and the UEs are randomly generated within the respective bounding boxes, and their $z$ coordinates are equal to $10$\,m, $10$\,m, and $1$\,m, respectively.
The RIS and target $z$ coordinates are equal to $15$\,m and $3$\,m.
Each AP power budget is $P=2$\,dBW,  with a bandwidth of $B=1$\,MHz, the noise power is $-204+10\log_{10}(B)$\,dBW. The channel's path loss follows the 3GPP urban microcell model \cite{3gpp2016technical}.
The clutter covariance matrix $\RR_r$ is modeled according to the local scattering model \cite{demir2022channel} with six clusters around$\oomega_{0,r}^{\textrm{RX}} \pm [20^\circ$, $10^\circ]$.
We have $T=5$ transmit APs and $R=2$ receiving APs, each equipped with $M=36$ antennas and serving $K=5$ UEs. We guarantee an SINR of $\gamma_k=3$\,dB to all UEs.
Unless otherwise specified, the RIS is equipped with $N=64$ elements, $C_1=C_2=C_3=C_4=2$, and the angles are uniformly distributed within an angle spread of $10^\circ$ in azimuth and elevation around the LoS angles between transmitter and receiver.
The results shown hereafter are averaged over $10$ random realizations of the AP positions and  UE positions.
For each of these realizations, the probability of detection is computed over $\tau=5$ timeslots and $1000\tau$ noise and clutter realizations. The false alarm probability is set to $10^{-4}$.

\begin{figure}[t!]
\begin{center}
%
%
\definecolor{mycolor1}{rgb}{0.00000,0.44700,0.74100}%
\definecolor{mycolor2}{rgb}{0.85000,0.32500,0.09800}%
\definecolor{mycolor3}{rgb}{0.92900,0.69400,0.12500}%
\definecolor{mycolor4}{rgb}{0.49400,0.18400,0.55600}%
\definecolor{mycolor5}{rgb}{0.46600,0.67400,0.18800}%
\begin{tikzpicture}

\begin{axis}[%
width=0.4\textwidth,
height=1.593in,
at={(0.758in,0.499in)},
scale only axis,
xmin=-40,
xmax=20,
xlabel style={font=\color{white!15!black}},
xlabel={$\delta^2$ [dB]},
ymin=0,
ymax=1,
legend columns=1,
ylabel style={font=\color{white!15!black}},
ylabel={$\text{P}_\text{d}$},
axis background/.style={fill=white},
axis x line*=bottom,
axis y line*=left,
xmajorgrids,
ymajorgrids,
legend style={nodes={scale=0.8, transform shape},at={(1,0.8)},legend cell align=left, align=left, draw=white!15!black}
]
\addplot [color=mycolor1, line width=2.0pt, mark=o, mark options={solid, mycolor1,mark size=3pt}]
  table[row sep=crcr]{%
-40	0.00079\\
-35	0.01702\\
-30	0.31095\\
-25	0.73967\\
-20	0.83065\\
-15	0.90721\\
-10	0.98953\\
-5	1\\
0	1\\
5	1\\
10	1\\
15	1\\
20	1\\
};
\addlegendentry{cl. aware}

\addplot [color=mycolor2, line width=2.0pt, mark=square, mark options={solid, mycolor2,mark size=3pt}]
  table[row sep=crcr]{%
-40	0.00062\\
-35	0.00412\\
-30	0.05406\\
-25	0.49787\\
-20	0.75228\\
-15	0.80024\\
-10	0.80304\\
-5	0.86958\\
0	0.95223\\
5	1\\
10	1\\
15	1\\
20	1\\
};
\addlegendentry{cl. unaware}

\addplot [color=mycolor3, dashed, line width=2.0pt, mark=o, mark options={solid, mycolor3,mark size=3pt}]
  table[row sep=crcr]{%
-40	0.00044\\
-35	0.019\\
-30	0.36604\\
-25	0.6711\\
-20	0.75984\\
-15	0.8578\\
-10	0.91823\\
-5	0.99874\\
0	1\\
5	1\\
10	1\\
15	1\\
20	1\\
};
\addlegendentry{aware, $C_{234}=20$}

\addplot [color=mycolor4, dashed, line width=2.0pt, mark=square, mark options={solid, mycolor4,mark size=3pt}]
  table[row sep=crcr]{%
-40	0.00068\\
-35	0.00475\\
-30	0.08446\\
-25	0.55357\\
-20	0.64647\\
-15	0.76255\\
-10	0.8015\\
-5	0.82778\\
0	0.9047\\
5	0.97948\\
10	1\\
15	1\\
20	1\\
};
\addlegendentry{unaware, $C_{234}=20$}

\addplot [color=mycolor5, dashed, line width=2.0pt, mark=triangle, mark options={solid, mycolor5,mark size=3pt}]
  table[row sep=crcr]{%
-40	0.00013\\
-35	0.00012\\
-30	0.00012\\
-25	0.00013\\
-20	0.00017\\
-15	0.00026\\
-10	0.00099\\
-5	0.00973\\
0	0.13111\\
5	0.36683\\
10	0.52755\\
15	0.97125\\
20	1\\
};
\addlegendentry{aware, random}

\end{axis}
\end{tikzpicture}%
      \vspace{-5mm}
	 \caption{Probability of detection vs RCS power with  CNR = $20$\,dB.}\label{pd 20}
		\end{center}
  \vspace{-5mm}
\end{figure}
\vspace{-3mm}
\begin{figure}[t!]
\begin{center}
%
%
\definecolor{mycolor1}{rgb}{0.00000,0.44700,0.74100}%
\definecolor{mycolor2}{rgb}{0.85000,0.32500,0.09800}%
\definecolor{mycolor3}{rgb}{0.92900,0.69400,0.12500}%
\definecolor{mycolor4}{rgb}{0.49400,0.18400,0.55600}%
\definecolor{mycolor5}{rgb}{0.46600,0.67400,0.18800}%
\begin{tikzpicture}

\begin{axis}[%
width=0.42\textwidth,
height=1.593in,
at={(0.758in,0.499in)},
scale only axis,
xmin=-40,
xmax=30,
xlabel style={font=\color{white!15!black}},
xlabel={$\delta^2$[dB]},
ymin=0,
ymax=1,
legend columns=2,
ylabel style={font=\color{white!15!black}},
ylabel={$\text{P}_\text{d}$},
axis background/.style={fill=white},
axis x line*=bottom,
axis y line*=left,
xmajorgrids,
ymajorgrids,
legend style={nodes={scale=0.8, transform shape},at={(1,0.7)},legend cell align=left, align=left, draw=white!15!black}
]
\addplot [color=mycolor1, line width=2.0pt, mark=o, mark options={solid, mycolor1,mark size=3pt}]
  table[row sep=crcr]{%
-40	0.00062\\
-35	0.0068\\
-30	0.18148\\
-25	0.65711\\
-20	0.79873\\
-15	0.80411\\
-10	0.89209\\
-5	0.99941\\
0	1\\
5	1\\
10	1\\
15	1\\
20	1\\
};
\addlegendentry{aware}

\addplot [color=mycolor2, line width=2.0pt, mark=square, mark options={solid, mycolor2,mark size=3pt}]
  table[row sep=crcr]{%
-40	0.00017\\
-35	0.00017\\
-30	0.00012\\
-25	0.00032\\
-20	0.00042\\
-15	0.00776\\
-10	0.22125\\
-5	0.54037\\
0	0.77072\\
5	0.8002\\
10	0.80197\\
15	0.86301\\
20	0.95411\\
};
\addlegendentry{unaware}

\addplot [color=mycolor3, dashed, line width=2.0pt, mark=o, mark options={solid, mycolor3,mark size=3pt}]
  table[row sep=crcr]{%
-40	0.00044\\
-35	0.00772\\
-30	0.25374\\
-25	0.63822\\
-20	0.70534\\
-15	0.74724\\
-10	0.81378\\
-5	0.93409\\
0	0.99999\\
5	1\\
10	1\\
15	1\\
20	1\\
};
\addlegendentry{aware, $C_{2,3,4}=20$}

\addplot [color=mycolor4, dashed, line width=2.0pt, mark=square, mark options={solid, mycolor4,mark size=3pt}]
  table[row sep=crcr]{%
-40	0.00016\\
-35	0.00021\\
-30	0.00011\\
-25	0.00021\\
-20	0.00075\\
-15	0.01443\\
-10	0.33281\\
-5	0.56467\\
0	0.65609\\
5	0.77981\\
10	0.80111\\
15	0.82398\\
20	0.90467\\
};
\addlegendentry{unaware, $C_{2,3,4}=20$}

\addplot [color=mycolor5, dashed, line width=2.0pt, mark=triangle, mark options={solid, mycolor5,mark size=3pt}]
  table[row sep=crcr]{%
-40	0.00011\\
-35	0.00015\\
-30	0.00017\\
-25	0.00017\\
-20	0.00021\\
-15	0.00027\\
-10	0.00056\\
-5	0.00523\\
0	0.07217\\
5	0.22387\\
10	0.4148\\
15	0.92222\\
20	1\\
};
\addlegendentry{ aware, random}

\end{axis}
\end{tikzpicture}%
      \vspace{-7mm}
	 \caption{Probability of detectionvs the RCS power with a CNR=$40$\,dB.}\label{pd 30}
		\end{center}
  \vspace{-4mm}
\end{figure}

\subsection{Clutter awareness and  cluster number's impact }
Fig. \ref{pd 20} shows this probability of detection vs.\ the RCS power, where we assumed that $\delta_{1,1}^2=\dots = \delta_{T,R}^2=\delta^2$ for a CNR of $20$\,dB. We can see that a clutter-aware system outperforms a clutter-unaware one: The clutter-unaware detector is defined as $\TT[\iota]=\sigma^{-2}\mathbf{I}_{MR}-
    \boldsymbol{\Xi}[\iota]$.
When $C_2=C_3=C_4=20$, denoted by $C_{234}$, the LoS assumption is no longer realistic, thus generating a performance decrease.
Our optimized model greatly outperforms the benchmark approach, consisting of a random RIS phase-shift matrix and an equal power allocation between the beams.
Fig. \ref{pd 30} shows what happens when the CNR rises to $40$\,dB. We see that clutter awareness rises in importance, widening the gap between the clutter-aware and unaware detectors.
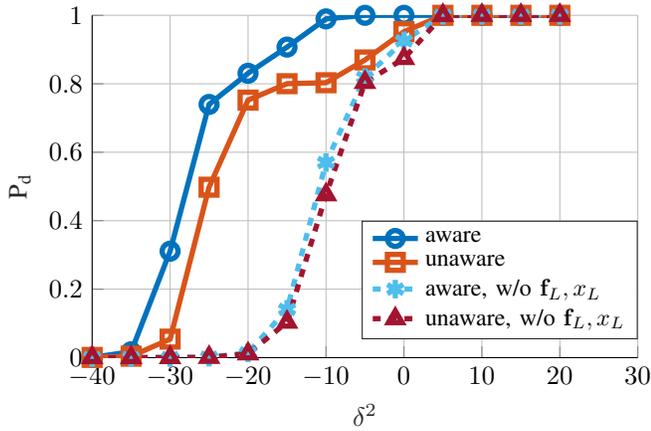
\begin{figure}[t!]
\begin{center}
%
%
\definecolor{mycolor1}{rgb}{0.00000,0.44700,0.74100}%
\definecolor{mycolor2}{rgb}{0.85000,0.32500,0.09800}%
\definecolor{mycolor3}{rgb}{0.3010 0.7450 0.9330}%
\definecolor{mycolor4}{rgb}{0.6350 0.0780 0.1840}%
\begin{tikzpicture}

\begin{axis}[%
width=0.40\textwidth,
height=1.593in,
at={(0.758in,0.499in)},
scale only axis,
xmin=-40,
xmax=30,
xlabel style={font=\color{white!15!black}},
xlabel={$\delta^2$},
ymin=0,
ymax=1,
ylabel style={font=\color{white!15!black}},
ylabel={$\text{P}_\text{d}$},
axis background/.style={fill=white},
axis x line*=bottom,
axis y line*=left,
xmajorgrids,
ymajorgrids,
legend style={nodes={scale=0.9, transform shape},at={(1,0.4)},legend cell align=left, align=left, draw=white!15!black}
]
\addplot [color=mycolor1, line width=2.0pt, mark=o, mark options={solid, mycolor1, mark size=3pt}]
  table[row sep=crcr]{%
-40	0.00079\\
-35	0.01702\\
-30	0.31095\\
-25	0.73967\\
-20	0.83065\\
-15	0.90721\\
-10	0.98953\\
-5	1\\
0	1\\
5	1\\
10	1\\
15	1\\
20	1\\
};
\addlegendentry{aware}

\addplot [color=mycolor2, line width=2.0pt, mark=square, mark options={solid, mycolor2,mark size=3pt}]
  table[row sep=crcr]{%
-40	0.00062\\
-35	0.00412\\
-30	0.05406\\
-25	0.49787\\
-20	0.75228\\
-15	0.80024\\
-10	0.80304\\
-5	0.86958\\
0	0.95223\\
5	1\\
10	1\\
15	1\\
20	1\\
};
\addlegendentry{unaware}

\addplot [color=mycolor3, dashed, line width=2.0pt, mark=asterisk, mark options={solid, mycolor3,mark size=3.5pt}]
  table[row sep=crcr]{%
-40	0.00017\\
-35	0.00021\\
-30	0.00032\\
-25	0.0013\\
-20	0.0141\\
-15	0.14094\\
-10	0.56988\\
-5	0.81853\\
0	0.92544\\
5	0.99996\\
10	1\\
15	1\\
20	1\\
};
\addlegendentry{aware, w/o $\ff_{L},x_{L}$}

\addplot [color=mycolor4, dashed, line width=2.0pt, mark=triangle, mark options={solid, mycolor4, mark size=3pt}]
  table[row sep=crcr]{%
-40	0.00018\\
-35	0.00019\\
-30	0.00024\\
-25	0.00082\\
-20	0.01112\\
-15	0.10365\\
-10	0.47577\\
-5	0.80401\\
0	0.87393\\
5	0.99966\\
10	1\\
15	1\\
20	1\\
};
\addlegendentry{unaware, w/o $\ff_{L},x_{L}$}

\end{axis}
\end{tikzpicture}%
      \vspace{-8mm}
	 \caption{Impact of a dedicated sensing stream onto the probability of detection with a CNR=$20$\,dB.}\label{sens vs no sens}
		\end{center}
  \vspace{-5mm}
\end{figure}

\subsection{Do we need dedicated sensing symbols?}
We have analyzed a system without sensing-dedicated precoding. Fig. \ref{sens vs no sens} compares the detection performance of systems with and without dedicated sensing precoding, both in the presence and absence of clutter awareness.
We can immediately see that removing the sensing streams implies a performance degradation.
Interestingly, the system without sensing dedicated precoding seems to suffer less from the lack of clutter awareness at a low RCS value. 

\section{Conclusions}
We have investigated a cell-free massive MIMO ISAC network jointly serving multiple communication UEs while detecting a potential target through an RIS in a multi-static fashion.
To this end, the transmit power allocation and the RIS phase-shift matrix are optimized to increase the target detection probability.
Numerical results show that clutter awareness plays a crucial role in the detection performance and that including a dedicated sensing stream benefits the system. The mmWave channel structure allows the system to assume LoS channels and still achieve good performances when the channel is composed of a high number of clusters concentrated in a small angular range.

 \bibliographystyle{IEEEtran}
\bibliography{IEEEabrv,auxiliary/biblio}
\end{document}